\documentclass[aps,preprint,superscriptaddress,amsmath,showpacs,longbibliography]{revtex4-1}
\usepackage{amssymb}
\usepackage{amsmath}
\usepackage{dcolumn}
\usepackage{graphicx}
\usepackage{mathrsfs}
\usepackage{appendix}
\usepackage{graphicx}
\usepackage{booktabs}
\usepackage{color}
\usepackage{mathrsfs}
\usepackage{bm}

\def \hide#1{}

\usepackage{url}
\usepackage[colorlinks]{hyperref}
\hypersetup{%
    plainpages=true,
    breaklinks=true,
    hypertexnames=false,
    pageanchor=true,
    colorlinks=true,
    linkcolor={blue},
    citecolor={red},
    urlcolor={blue},
    anchorcolor={black}
}

\begin{document}
\title{Generating Synthetic Magnetism via Floquet Engineering Auxiliary Qubits in Phonon-Cavity-Based Lattice}

\author{Xin Wang}
\affiliation{Institute of Quantum Optics and Quantum Information,
	School of Science, Xi'an Jiaotong University, Xi'an 710049, China}
\affiliation{MOE Key Laboratory for Nonequilibrium Synthesis and
	Modulation of Condensed Matter, School of Science, Xi'an Jiaotong University, Xi'an 710049, China}

\author{Hong-Rong Li}
\email{hrli@mail.xjtu.edu.cn}
\affiliation{Institute of Quantum Optics and Quantum Information,
	School of Science, Xi'an Jiaotong University, Xi'an 710049, China}
\affiliation{MOE Key Laboratory for Nonequilibrium Synthesis and
	Modulation of Condensed Matter, School of Science, Xi'an Jiaotong University, Xi'an 710049, China}

\author{Fu-Li Li}
\affiliation{Institute of Quantum Optics and Quantum Information,
	School of Science, Xi'an Jiaotong University, Xi'an 710049, China}
\affiliation{MOE Key Laboratory for Nonequilibrium Synthesis and
	Modulation of Condensed Matter, School of Science, Xi'an Jiaotong University, Xi'an 710049, China}

\email{hrli@mail.xjtu.edu.cn}

\begin{abstract}
Gauge magnetic fields have a close relation to breaking time-reversal symmetry in condensed matter. In the present of the gauge fields, we might observe nonreciprocal and topological transport. Inspired by these, there is a growing effort to realize exotic transport phenomena in optical and acoustic systems. However, due to charge neutrality, realizing analog magnetic flux for phonons in nanoscale systems is still challenging in both theoretical and experimental studies.
Here we propose a novel mechanism to generate synthetic magnetic field for phonon lattice by Floquet engineering auxiliary qubits. We find that, a longitudinal Floquet drive on the qubit will produce a resonant coupling between two detuned acoustic cavities. Specially, the phase encoded into the longitudinal drive can exactly be transformed into the phonon-phonon hopping. Our proposal is general and can be realized in various types of artificial hybrid quantum systems. Moreover, by taking surface-acoustic-wave (SAW) cavities for example, we propose how to generate synthetic magnetic flux for phonon transport. In the present of synthetic magnetic flux, the time-reversal symmetry will be broken, which allows to realize the circulator transport and analog Aharonov-Bohm effects for acoustic waves. Last, we demonstrate that our proposal can be scaled to simulate topological states of matter in quantum acoustodynamics system.	
\bigskip

keywords: synthetic magnetism; Floquet engineering; quantum simulations; quantum acoustodynamics
\end{abstract}

\maketitle

\section{introduction} 
A moving electron can feel a magnetic field, and due to the path-dependent phases, many exotic transport phenomena, including the Aharonov-Bohm (AB) effect, matter-wave interference and chiral propagation, can be observed in experiments~\cite{Aharonov1959,vanOudenaarden1998,Vidal2000}. In topological  matters, an external magnetic field is often used to break time reversal symmetry (TRS) of the whole system. For example, in the present of magnetic fields, the integer quantum Hall effect allows to observe topologically protected unidirectionally transport around the edge~\cite{Klitzing1980,MacDonald1984,Hatsugai1993}.  In photonic and phononic systems, mimicking the dynamics of charged particles allows to observe the phenomena of topological matters and realizing quantum simulations of strongly correlated systems~\cite{Dalibard11,WangP2015,Fleury2016}. By breaking TRS, the chiral and helical transports for classical acoustic waves were already demonstrated in macroscopic meta-material setups such as coupled spring systems and circulating fluids~\cite{Fleury2014,Nash2015,Yang15,Fleury2016,Ssstrunk2016}.

However, due to the charge neutrality, phonons or photons, cannot acquire the vector-potential induced phase (analog to Peierls phases) effectively~\cite{Fang2012,Peano2015}. To this end, one needs to create analog magnetic fields via specific control methods. Much efforts have been devoted into 
realizing synthetic magnetism for neutral particles, in the artificial quantum systems~\cite{Dalibard11,Estep2014,Schmidt2015,Roushan2016,Fang2017,Peterson2017,Ozawa2019}. These include exploring topological photonics and phononics in circuit-QED-based photon lattices, optomechanical arrays and microwave cavity systems~\cite{Hafezi2011,Bermudez2011,Mittal2014,Peano2015,Deymier2018,Mittal2019}. 
For example, as discussed in Ref.~\cite{Schmidt2015,Mathew2018}, one can artificially generate synthetic magnetism for photons in an optomechanical systems via site-dependent modulated laser fields. 

In recent years, manipulating acoustic waves at a single-phonon level has drawn a lot of attention~\cite{Viennot2018}, and the quantum acoustodynamics (QAD) based on piezoelectric surface acoustic waves (SAW) has emerged as a versatile platform to explore quantum features of acoustic waves~\cite{Schuetz15,Manenti16,Manenti2017,Moores2018,Satzinger2018,Kockum2018,Shao2019,Sletten2019}. Unlike localized mechanical oscillations~\cite{Blencowe2004,Meystre2012,Aspelmeyer14}, the piezoelectric surface is much larger compared with the acoustic wavelength. Therefore, the phonons are itinerant on the material surface and in the form of propagating waves~\cite{Manenti2017}.
To control the quantized motions effectively, the SAW cavities are often combined with other systems, for example, with superconducting qubits or color centers, to form as hybrid quantum systems~\cite{Delsing2019}.
Analog to quantum controls with photons,  many acoustic devices, such as circulators, switches and routers, are also needed for generating, controlling, and detecting SAW phonons in QAD experiment~\cite{Kittel1958,Sun2012,Habraken2012,Ekstrm2017,Fleury2014,Khanikaev2015,He2016,Wang2018,Shao2019}.
However, it is still challenging to realize these nonreciprocal acoustic quantum devices in the artificial nano- and micro-scale SAW systems~\cite{Brendel2017,Mathew2018}.

By applying an external time periodic drive on a static system, the effective Hamiltonian of the whole system for the longtime evolution can be tailed freely, which allows to observe topological band properties. This approach is referred as Floquet engineering~\cite{Goldman2014,Goldman2015,Vogl2019}, and has been exploited to simulate the topological toy models such as 
Haldane and Harper-Hofstadter lattices in artificial systems~\cite{Ozawa2019}.
Inspired by these studies, we propose a novel method to realize synthetic magnetism in the phonon systems by Floquet engineering qubits coupled to phonons. Different from previous methods based on optomechanical systems~\cite{Peano2015,Peterson2017,Mathew2018}, we find that the Floquet driving can be effectively realized in an opposite way, i.e, via the unconventional cavity optomechanics (UOM) discussed in Ref.~\cite{Wang19}, where the mechanical frequency is effectively modulated by an electromagnetic field via an auxiliary qubit. We find that, the phase of the longitudinal drive on the qubit, can be effectively encoded into the phonon transport between two SAW cavities. Consequently, considering a SAW-based phonon lattice with closed loops, the phonons can feel a synthetic magnetic flux and behave as charged particles.

In our discussions, we take the SAW-based phonon lattice as an example. However, the proposed method to generate synthetic magnetism is general and can be applied to other artificial quantum systems, for realizing nonreciprocal control of phonons, or even photons.
We show that, the TRS will be broken by tuning the synthetic magnetism flux of a loop in the lattice, and it is possible to realize a phonon circulator and simulate Aharonov-Bohm effect in an acoustic system. 
Additionally, our findings might pave another way to simulate the topological models by breaking TRS for charge neutral polaritons. 

\section{qubit-intermediated SAW-cavity coupling}
\begin{figure}[tbph]
	\centering \includegraphics[width=8.2cm]{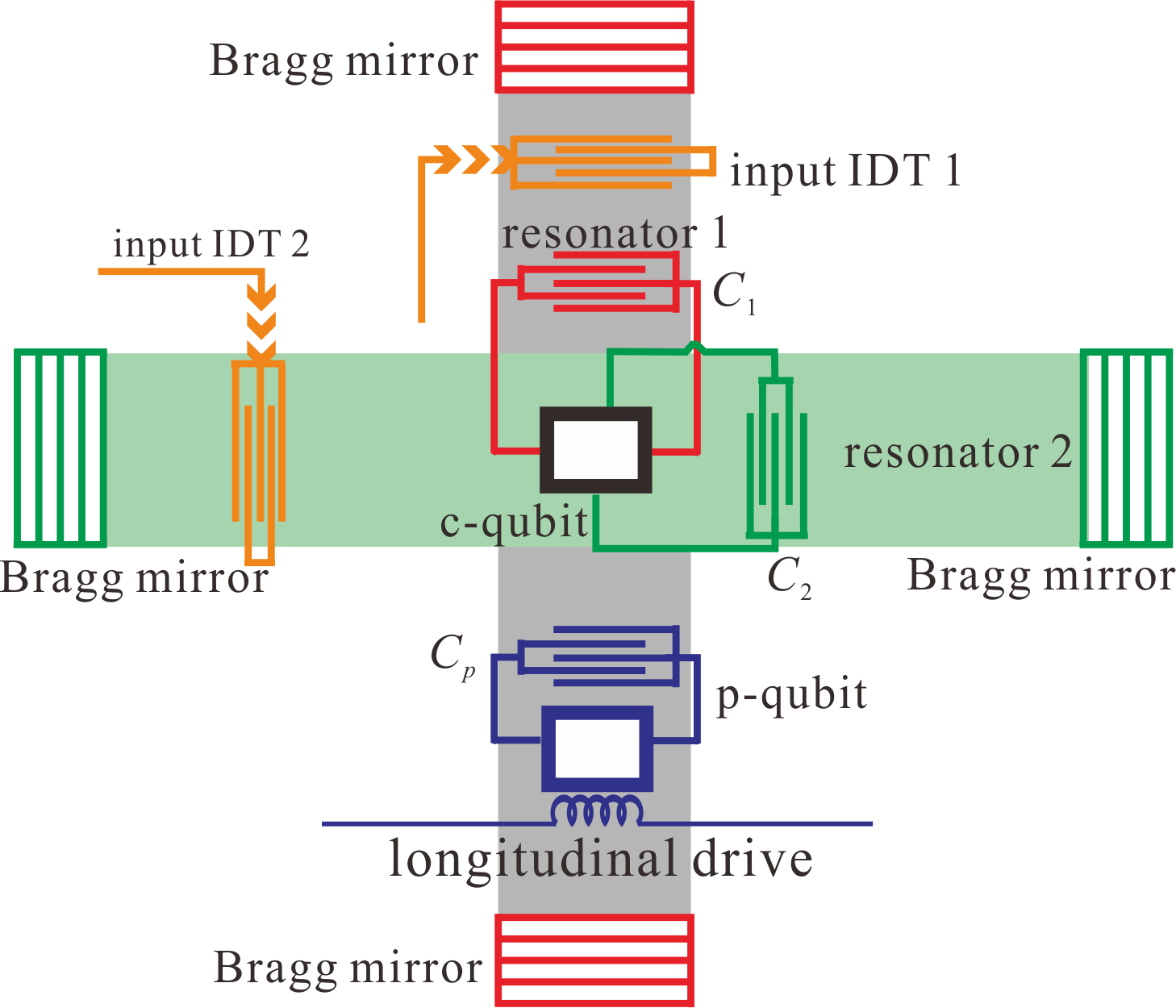}
	\caption{The hopping model of two surface acoustic wave (SAW) cavities for generating synthetic magnetisms via Floquet engineering: the SAW cavity 1 (grey) and 2 (green) are mediated via a superconducting qubit (defined as c-qubit) located at their cross section. The qubit-SAW coupling is mediated via the capacitances $C_{1,2}$ of interdigitated-transducer (IDT) type (via the piezoelectric effect). To realize time-dependent Floquet engineering, the cavity 1 is dispersively coupled with another qubit (noted as p-qubit) via capacitance $C_{p}$. The p-qubit is longitudinally driven by a classical electromagnetic field, which will effectively change the resonant frequency of the phonon cavity 1 periodically.}
	\label{fig1m}
\end{figure}

We start our discussions by considering two coupled mechanical modes. The coupling can be direct, or intermediated by auxiliary element such as qubit or photon cavities. For example,
as shown in Fig.~\ref{fig1m}, we assume that two SAW cavities (confined by two phonon Bragg mirrors), are coupled to a same superconducting transmon (or charge) qubit located at the cross section~\cite{Makhlin2001,Irish03,Koch071,Gu2017}. The couplings are mediated by capacitances of the interdigitated-transducer (IDTs) type~\cite{Manenti16,Satzinger2018}. In our discussions, the qubit (noted as c-qubit) is employed for realizing a tunable couplings between two acoustic modes. The acoustic motions on piezoelectric crystal surface will produce phonon-induced oscillating voltages on the interdigitated capacitances $C_{i}$~\cite{Manenti16,Manenti2017}.
Note that the frequency separations between the discrete acoustic modes are much narrower than those in the optical cavities, which might be one obstacle for single-mode quantum control~\cite{Moores2018}. In experiments, one can effectively select one resonant mode coupled to the qubit by: (i) designing the Bragg mirror to achieve high reflection for a narrow bands~\cite{Satzinger2018}, (ii) controlling positions of the IDT electrodes, and (iii) setting the space periodicity of IDT capacitance the same with the coupled SAW modes~\cite{Manenti2017,Bolgar18}. Those methods allow to select a single SAW mode to couple with the superconducting qubit, and the qunatum control and measurement of a single acoustic phonon has been experimentally realized~\cite{Bolgar18,Manenti2017,Satzinger2018,Ask2019}.
Consequently, the Hamiltonian for the system in Fig.~\ref{fig1m} reads
\begin{equation}
H_{12}=\frac{\omega_{qc}}{2}\sigma_{c}^{z}+\sum_{i=1,2}\omega_{mi}b_{i}^{\dagger}b_{i}
+\sum_{i=1,2}g_{i}(\sigma_{c}^{+}b_{i}+\sigma_{c}^{-}b_{i}^{\dagger}),
\end{equation}
where and $b_{i}$ and ($b_{i}^{\dagger}$) is the annihilation (creation) operators of $i$th SAW cavity, respectively; $\sigma_{c}^{z}=|e_{c}\rangle\langle e_{c}|-|g_{c}\rangle\langle g_{c}|$ and $\sigma_{c}^{x}=|e_{c}\rangle\langle g_{c}|+|g_{c}\rangle\langle e_{c}|$ are the qubit Pauli operators, where $|g_{c}\rangle$ and $|e_{c}\rangle$ are the ground and excited states of the c-qubit with $\omega_{qc}$ being the transition frequency.

In experiments, the induced dispersive coupling between a qubit and a SAW mode based on large detuning has been observed~\cite{Manenti2017,Sletten2019}. We assume that the two cavities are of large detuning with the c-qubit (i.e., in the dispersive coupling regime)~\cite{Bolgar18,Sletten2019}. In the rotating frame, the Hamiltonian $H_{12}$ becomes
	\begin{equation}
	H_{12}(t)=g_{1}(\sigma_{c}^{+}b_{1}e^{i\Delta_{1}t}+\sigma_{c}^{-}b_{1}^{\dagger}e^{-i\Delta_{1}t})+g_{2}(\sigma_{c}^{+}b_{2}e^{i\Delta_{2}t}+\sigma_{c}^{-}b_{2}^{\dagger}e^{-i\Delta_{2}t}),
	\end{equation}
	where $\Delta_{i}=\omega_{qc}-\omega_{mi}$ is the qubit-SAW detuning. Given that the considered time scale is much larger than $(\Delta_{1,2})^{-1}$, we can write the effective Hamiltonian as~\cite{James2007} 
	\begin{equation}
	H_{12}(t)=\sum_{m,n=1}^{2}\frac{1}{2}(\frac{1}{\Delta_m}+\frac{1}{\Delta_n})[g_{m}\sigma_{c}^{+}b_{m}, g_{n}\sigma_{c}^{-}b^{+}_{n}]e^{i(\Delta_{m}-\Delta_{n})t},
	\end{equation}
	In our discussion, we assumed that $|\Delta_{2}-\Delta_{1}| \ll \{|\Delta_{1}|,|\Delta_{2}|\}$. Therefore, $H_{12}(t)$ can be rewritten as
	\begin{equation}
	H_{12}(t)=\sum_{i=1,2} \frac{g_{i}^{2}}{\Delta_i}\sigma_{c}^{z}(b^{\dagger}_{i}b_{i}+\frac{1}{2})+\frac{g_{1}g_{2}}{2}(\frac{1}{\Delta_2}+\frac{1}{\Delta_1})\left[b_{1}^{\dagger}b_{2}e^{i(\Delta_{2}-\Delta_{1})t}+\text{H.c.}\right],
	\end{equation}
	where the first term will produce Stark shifts on SAW cavity and Lamb shift of the qubit.
	One can eliminate the qubit degree of freedom by assuming the c-qubit approximately in its ground state with $\sigma_{c}^{z}\simeq-1$. For simplification, we assume $g=g_{1}=g_{2}$, and obtain
	\begin{equation}
	H_{12}(t)=-\sum_{i=1,2} \frac{g^{2}}{\Delta_i}b^{\dagger}_{i}b_{i}+\frac{g^2}{2}(\frac{1}{\Delta_2}+\frac{1}{\Delta_1})\left[b_{1}^{\dagger}b_{2}e^{i(\Delta_{2}-\Delta_{1})t}+\text{H.c.}\right].
	\label{H_mediate}
	\end{equation}
	 In the rotating frame of the first term in Eq.~(\ref{H_mediate}), we obtain the interacting Hamiltonian between two phonon cavities mediated by the c-qubit
	\begin{equation}
	H_{c12}(t)=g_{12}(b_{1}^{\dagger}b_{2}e^{i\delta'_{12}t}+\text{H.c.}). \quad g_{12}=\frac{1}{2}\sum_{i=1,2}\frac{g^2}{\Delta_{i}},  \quad \delta'_{12}=(\Delta_{2}-\Delta_{1})(1-\frac{g^2}{\Delta_{1}\Delta_{2}}),
	\end{equation}
	from which we find that the detuning between two cavities is slightly shifted from $(\Delta_{2}-\Delta_{1})$. In the limit $g^2\ll |\Delta_{1}\Delta_{2}|$, we can approximately obtain $\delta'_{12}\simeq(\Delta_{2}-\Delta_{1})$.
	
It should be noted that generating phonon-cavity coupling mediated by c-qubits is not a necessary step for our proposal. We discuss this since an intermediate coupling allows more controlling flexibility. Alternatively, as discussed in Refs.~\cite{Brown2011,Okamoto2013}, one can directly couple different mechanical modes without other auxiliary qunatum elements. Our following discussions can also be applied to the phonon-cavity lattice based on the direct couplings.

\section{Generating synthetic magnetism for phonons}
To encode a phase factor into the phonon transport between SAW cavities, we consider SAW cavity 1 is coupled to another p-qubit (the transition frequency being $\omega_{p}$) with strength $g_{p}$. 
The p-qubit is longitudinally driven by an external field~\cite{Zhao15,Billangeon2015,Richer16,Richer17,Goetz18}, i.e., $$H_{d}=-\frac{\Omega_{p}}{2}  \sigma_{p}^{z}\cos(\omega_{d}t+\phi),$$ 
with $\phi$ being the relative phase. In the rotating frame at the frequency $\omega_{m1}$,
the Hamiltonian for the Floquet engineering process reads
\begin{equation}
H_{p}(t)=-\frac{\Delta_{p}}{2}\sigma_{p}^{z}+g_{p}(\sigma_{p}^{-}b_{1}^{\dagger}+\sigma_{p}^{+}b_{1})-\frac{\Omega_{p}}{2}\sigma_{p}^{z}\cos(\omega_{d}t+\phi).
\label{hp}
\end{equation}
where $\Delta_{p}$ is the detuning between the p-qubit and cavity 1. 
The drive $H_{d}$ is of longitudinal form and on the qubit operator $\sigma_{z}$ (rather than the conventional dipole drive on qubit operator $\sigma_{x}$). Therefore $H_{d}$ commutes with the qubit frequency operator. In the rotating frame, the driving frequency $\omega_d$ is kept unchanged.

In the limit $\Omega_{p}\rightarrow0$ and under the condition $\Delta_{p}\gg g_{p}$, $H_{p}$ results in a standard dispersive coupling~\cite{Gu2017}
\begin{equation}
H_{\text{dis}}=-\chi_{0}\sigma_{p}^{z}b_{1}^{\dagger}b_{1}, \qquad \chi_{0}=\frac{g_{p}^{2}}{\Delta_{p}},
\label{chi_zz}
\end{equation}
where $H_{\text{dis}}$ is the standard dispersive coupling with $\chi_{0}$ being the coupling strength. Considering $\Omega_p$ to be nonzero and satisfying $\Omega_p \ll \Delta_p$, we can view the longitudinal drive $H_{d}$ as modulating the detuning $\Delta_{p}$ time-dependently, i.e.,
\begin{equation}
\Delta_{p}\longrightarrow \Delta_{p}(t)=\Delta_{p}+\Omega_{p}\cos(\omega_{d}t+\phi).
\end{equation}
During the modulation process, we assume that the system is always in the dispersive regime, which requires 
\begin{equation}
|\Delta_{p}(t)|\gg g_{p}.\label{discond}
\end{equation}
Similar to the discussions of UOM mechanisms in Ref.~\cite{Wang19}, one can replace the constant dispersive coupling as a time-dependent form, i.e., $\chi_{0}\rightarrow \chi(t)$. Since the coupling is always of large detuning, the phonon cannot excite the qubit effectively. We assume that the p-qubit is approximately in its initial ground state, i.e., $\sigma_{p}^{z}\simeq-1$, Therefore, the dispersive coupling in Eq.~(\ref{chi_zz}) can be viewed as a periodic modulation of the SAW cavity frequency, i.e., 
\begin{equation}
H_{\text{dis}}(t)=\chi(t)b_{1}^{\dagger}b_{1}, \quad \chi(t)=\frac{g_{p}^{2}}{\Delta_{p}+\Omega_{p}\cos(\omega_{d}t+\phi)},
\label{H_dis1}
\end{equation}
which is a time-periodic Hamiltonian and satisfies 
\begin{equation}
H_{\text{dis}}(t)=H_{\text{dis}}(t+T), \quad T=\frac{2\pi}{\omega_{d}}.
\end{equation}
\begin{figure}[tbph]
	\centering \includegraphics[width=8.8cm]{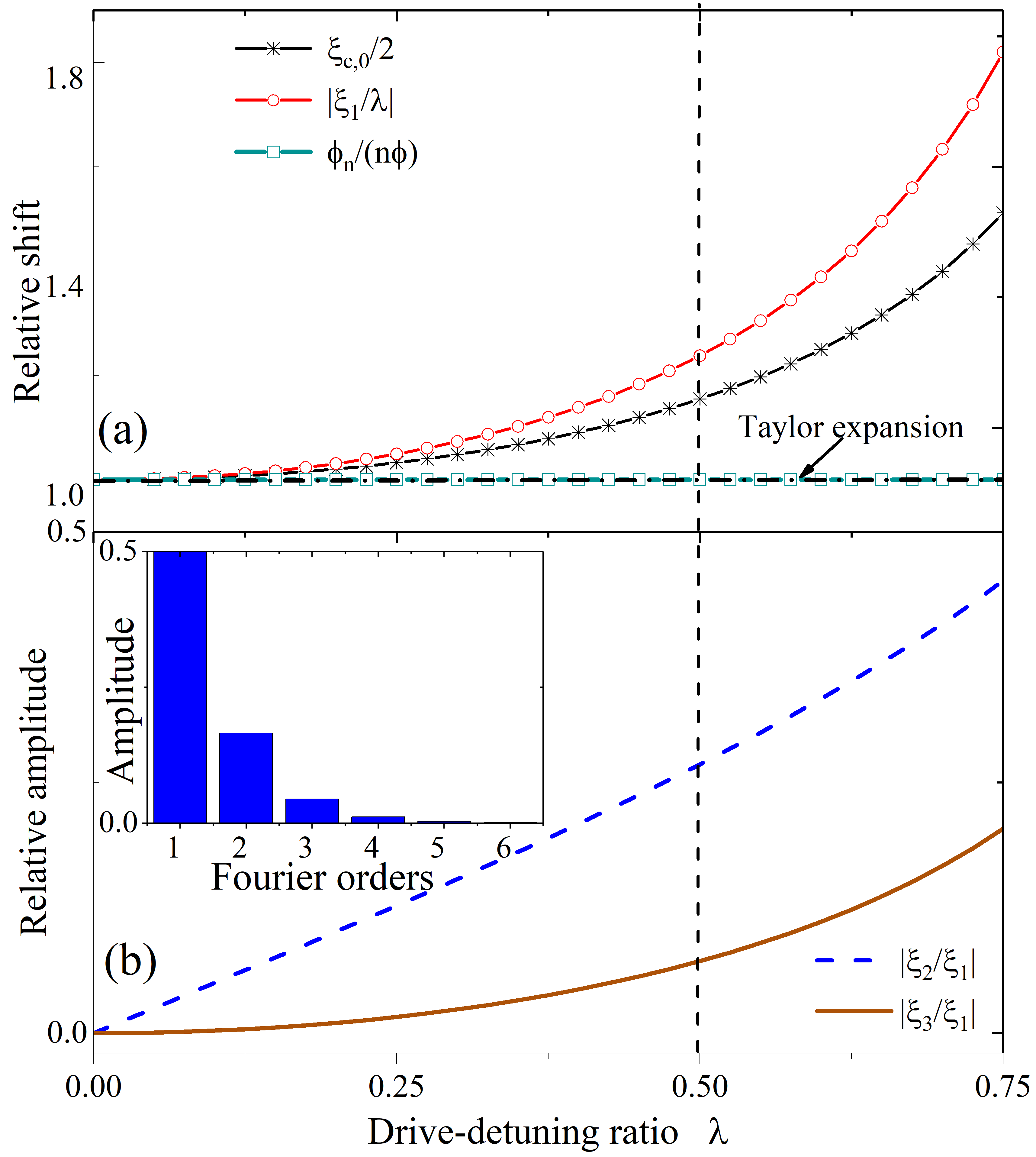}
	\caption{(a) The amplitude $\xi_{1,2}$ and phase $\phi_{1}$ for the zero- and first-order frequency components change with the dimensionless parameter $\lambda$. The dashed horizon line corresponds to the the Taylor expansion results.
	We find that the amplitudes $\xi_{1,2}(\lambda)$ are highly nonlinear functions of large $\lambda$. While for $n\leq 6$, the phase of the first-order $\phi_{n}$
	 is always equal to $n\phi$ (encoded into the Floquet engineering of the p-qubit). (b) The relative amplitude ratios $\xi_{2}/\xi_{1}$ and $\xi_{3}/\xi_{1}$ change with the longitudinal drive-detuning ratio $\lambda$. The insets displays the amplitudes changing with the Fourier orders at $\lambda=0.5$ (the dashed line position).}
	\label{fig2m}
\end{figure}

As discussed in Ref.~\cite{Eckardt2015}, a suitable time-periodic drive will lead an effective Hamiltonian, which alters the long-time dynamics significantly. This Floquet engineering method requires to obtain the frequency information of time-periodic drive first.
The Hamiltonian $H_{\text{dis}}(t)$ is nonlinear when the parameter $\lambda=\Omega_{p}/\Delta_{p}$ deviates from zero significantly. We expand $\chi(t)$ in the Fourier form:
\begin{subequations}
	\begin{gather}
	\chi(t)=\chi_{0}\left[\frac{\xi_{c,0}}{2}+\sum_{n=1}^{\infty}\xi_{n}\cos(n\omega_{d}t+\phi_{n})\right],\label{chitt}\\
	\xi_{n}=\sqrt{\xi_{c,n}^{2}+\xi_{s,n}^{2}}, \quad \phi_{n}=-\arctan{\frac{\xi_{s,n}}{\xi_{c,n}}},
	\end{gather}
\end{subequations}
where $\xi_{n}$ ($\phi_{n}$) is the amplitude (phase) of $n$-th order frequency component, and $\xi_{c(s),n}$ is defined as
\begin{subequations}
	\begin{gather}
	\xi_{c,n}=\frac{\omega_{d}}{\pi}\int_{-\pi/\omega_{d}}^{\pi/\omega_{d}}\frac{\cos(n\omega_{d}t)}{1+\lambda\cos(\omega_{d}t+\phi)}dt, \\
	\xi_{s,n}=\frac{\omega_{d}}{\pi}\int_{-\pi/\omega_{d}}^{\pi/\omega_{d}}\frac{\sin(n\omega_{d}t)}{1+\lambda\cos(\omega_{d}t+\phi)}dt.
	\end{gather}
	\label{xicn}
\end{subequations} 
In the limit $\lambda\rightarrow0$, $\chi(t)$ can be written in the Taylor expansion form (to the first-oder), and then
\begin{equation}
\frac{\xi_{c,0}}{2}\simeq 1, \quad \xi_{1}\simeq -\lambda, \quad \phi_{1}=\phi.
\label{appro}
\end{equation}
 
In Fig.~\ref{fig2m}(a), we numerically plot $\xi_{0}/2$, $|\xi_{1}/\lambda|$ and $\phi_{1}/\phi$ versus $\lambda$, respectively. The dashed horizontal line is at $\xi_{0}/2=|\xi_{1}/\lambda|=\phi_{1}/\phi=1$, which corresponds to the Taylor expansion results. We find that, both $\xi_{0}(\lambda)$ and $\xi_{1}(\lambda)$ are nonlinear functions of $\lambda$, and shift from Taylor expansion results significantly. In fact, by numerical Fourier transformation, we find that $\phi_{n} =n\phi$ is always valid for $n\leq6$ as shown in Fig.~2(a). This indicates that \emph{the phase information encoded in the longitudinal drive is exactly kept even for large $\lambda$}. 
This point is very important to control the synthetic magnetism in the acoustic loop, which will be discussed in detail in the following discussion.

In Fig.~\ref{fig2m}(b), we plot the relative amplitudes $\xi_{2,3}/\xi_{1}$ changes with $\lambda$. We find that, although $\xi_{2,3}/\xi_{1}$ increase with $\lambda$, $\xi_{2,3}/\xi_{1}<1$ is always valid even with a large $\lambda$. In the inset plot, we plot $\xi_{n}$ changing with $n$ by adopting $\lambda=0.5$, and find that $\xi_{n}$ will decrease quickly with increasing the Fourier harmonic order $n$. Note that the zero-order component, $\xi_{c,0}/2$, will produce a frequency shift of the frequency of the acoustic mode. 

In our discussion, we do not require $\lambda$ to be very small. However, $\lambda$ cannot be too close to one, since the dispersive condition in Eq.~(\ref{discond}) will not be valid. Additionally, the high-order frequency components might also produce observable effects. 
For $0<\lambda<1$, the effective Hamiltonian is taken in the form
\begin{eqnarray}
H_{12}^{(0)}(t)&=&H_{c12}(t)+H_{\text{dis}}(t)\notag\\
&=&g_{12}(b_{1}^{\dagger}b_{2}e^{i\delta_{12}t}+b_{2}^{\dagger}b_{1}e^{-i\delta_{12}t})+\chi_{0}\sum_{n=1}^{\infty}\xi_{n}\cos(n\omega_{d}t+n\phi)b_{1}^{\dagger}b_{1}, 
\label{H120t}
\end{eqnarray}
where
\begin{equation}
\delta_{12}=\delta'_{12}+\frac{2g_{12}^{2}}{\delta'_{12}}+\frac{\chi_{0}}{2}\xi_{c,0}
\label{detune12}
\end{equation}
is the renormalized frequency detuning. Note that we have already included the Stark shift $(2g_{12}^{2})/\delta'_{12}$ into $\delta_{12}$. The Floquet amplitudes satisfy $\xi_{n+1}\ll \xi_{n}$. The last part in $H_{12}^{(0)}(t)$ is periodic. As discussed in Ref.~\cite{Eckardt2015}, given that the considered time scale is much longer than $(\omega_{d})^{-1}$, we apply the unitary transformation to $H_{12}^{(0)}(t)$ [Eq.~(\ref{H120t})]:
	\begin{eqnarray}
	|\psi_{s}\rangle &=&U(t)|\psi_{s}^{0}\rangle, \quad U(t)=\prod_{n}\mathcal{R}_n(t), \notag \\
	H_{12}(t)&=&U(t)H_{12}^{(0)}(t)U^{\dagger}(t)-iU(t)\frac{\partial U^{\dagger}(t)}{\partial t}, 
	\label{hrot}
	\end{eqnarray}
where $|\psi_{s}^{0}\rangle$ ($|\psi_{s}\rangle$) is original (transformed) state of the system, $H_{12}(t)$ is the transformed effective Hamiltonian, and $\mathcal{R}_{n}(t)$ is the unitary rotation operator for $n$th order:
\begin{equation}
\mathcal{R}_{n}(t)=\exp\left[i K_{n} b_{1}^{\dagger}b_{1}\sin(n\omega_{d}t+\phi_{n})\right]=\sum_{k=-\infty}^{k=\infty}\mathcal{J}_{k}(K_{n}b_{1}^{\dagger}b_{1})e^{ik(n\omega_{d}t+\phi_{n})}, \quad K_{n}=\frac{\chi_{0}\xi_{n}}{n\omega_{d}},
\label{knn}
\end{equation}
where $\mathcal{J}_{k}$ is the $k$th-order Bessel function of the first kind, and $K_{n}$ is the dimensionless parameter~\cite{Goldman2014,Goldman2015}.

In our discussion, we set the $\chi_{0}/\omega_{d}<1$. As shown in Fig.~\ref{fig2m}(b), given that $\lambda<1$, one finds that $\xi_{n}<1$ and $\xi_{n}\ll \xi_{n-1}$. Therefore, for $n\geq1$, it is reasonable to assume that 
\begin{equation}
K_{1}\ll 1, \quad K_{n}\ll K_{n-1}.
\end{equation}
Moreover, we set the detuning between two cavities is around the first-order frequency, i.e., $|\delta_{12}|\simeq \omega_{d}$. All the higher order frequency components ($n\geqslant2$), are not only of weak strengths ( $K_{n}\ll1$), but also of large detunings. Therefore, we only consider the first-oder unitary rotation $\mathcal{R}_{1}(t)$, i.e., 
\begin{equation}
U(t)\simeq \mathcal{R}_{1}(t)=\sum_{k=-\infty}^{k=\infty}\mathcal{J}_{k}(K_{1}b_{1}^{\dagger}b_{1})e^{ik(\omega_{d}t+\phi)}.
\end{equation}

By deriving Eq.~(\ref{hrot}), one finds that 
	\begin{equation}
	U(t)H_{\text{dis}}(t)U^{\dagger}(t)-iU(t)\frac{\partial U^{\dagger}(t)}{\partial t}=0.
	\end{equation}	
	Therefore, 
	\begin{equation}
	H_{12}(t)=U(t)H_{c12}(t)U^{\dagger}(t).
	\end{equation}
	In the long-time limit $t\gg 1/\omega_{d}$, the Floquet-engineered effective Hamiltonian $H_{12}(t)$ is approximately written as
\begin{equation}
H_{12}(t)\backsimeq g_{12}\left[\mathcal{R}_{1}(t) b_{1}^{\dagger}b_{2}\mathcal{R}_{1}^{\dagger}(t)e^{i\delta_{12}t}+\text{H.c.}\right].
\label{oneorder}
\end{equation}
Under the condition $|K_1b^\dagger_1 b_1|\ll \sqrt{2}$, we can expand the Bessel function $\mathcal{J}_{k}(K_{1}b_{1}^{\dagger}b_{1})$ as
	\begin{equation}
	\mathcal{J}_{k}(K_{1}b_{1}^{\dagger}b_{1})=(-1)^{k}\mathcal{J}_{-k}(K_{1}b_{1}^{\dagger}b_{1})\backsimeq\frac{1}{k!}\left(\frac{K_{1}b_{1}^{\dagger}b_{1}}{2}\right)^{k},
	\end{equation}
	which is valid in the low-phonon-number limit with a small parameter $K_{1}<1$. In the following discussion, we focus on the cases where the intracavity phonon number is around the quantum regime. Under these conditions,
	We expand $\mathcal{R}_{1}(t)$ as
	\begin{equation}
	\mathcal{R}_{1}(t)\simeq \sum_{k=0}^{k=+\infty}C_{k}\left[e^{ik(\omega_{d}t+\phi)}+(-1)^{k}e^{-ik(\omega_{d}t+\phi)}\right](b_{1}^{\dagger}b_{1})^{k}, \quad
	C_{k}=\left\{
	\begin{array}{lr}
	\frac{1}{2}\quad k=0, &  \\
	\frac{1}{k!}\left(\frac{K_{1}}{2}\right)^{k} \quad k>0.  & 
	\end{array}
	\right.
	\end{equation}
	Now we consider the resonant modulation case with $\delta_{12}=-\omega_{d}$, and obtain the following effective Hamiltonian 
	\begin{eqnarray}
	H_{12}^{\text{eff}}&=&J_{12}\left\{\sum_{k=0}^{k=+\infty}C_{k}\left[e^{ik(\omega_{d}t+\phi)}+(-1)^{k}e^{-ik(\omega_{d}t+\phi)}\right](b_{1}^{\dagger}b_{1})^{k} \right\} b_{1}^{\dagger}b_{2}   \notag\\
	&&\times\left\{\sum_{k=0}^{k=+\infty}C_{k}\left[e^{-ik(\omega_{d}t+\phi)}+(-1)^{k}e^{ik(\omega_{d}t+\phi)}\right](b_{1}^{\dagger}b_{1})^{k} \right\}e^{-i\omega_d t}+\text{H.c.},
	\label{Heff}
	\end{eqnarray}
	where we can adopt the rotating wave approximation by neglecting the oscillating terms at frequency $k\omega_d$ ($|k|\geq1$) under the condition $|C_{k}J_{12}|<|\omega_d|$. Consequently, only multiplying the nearest orders with coefficients $C_{k}$ and $C_{k\pm1}$ will contribute to the resonant terms. Due to the relation
	$$2C_{0}\simeq 1\gg {C_1}>{C_{k}}, \quad k>1,$$
	we obtain 
	\begin{equation}
2C_{0}{C_1}\gg C_{1}{C_2}>C_{2}{C_3}... \label{Fcondition}
	\end{equation}
	Therefore, we just need to consider the zero- and first-order terms in Eq.~(\ref{Heff}). Consequently, we obtain the following effective Hamiltonian
	\begin{equation}
	H_{12}^{\text{eff}}=J_{12}e^{i\phi}b_{1}^{\dagger}b_{2}+\text{H.c.}, \quad J_{12}\simeq 2C_{0}C_{1}g_{12}=\frac{g_{12}K_{1}}{2},
	\label{Hefflong}
	\end{equation} 
where we find that the relative phase $\phi$ is successfully encoded into the coupling between cavity 1 and 2. Note that by choosing the opposite detuning relation $\omega_{d}=\delta_{12}$, the encoded phase becomes $-\phi$. 
\begin{figure}[tbph]
	\centering \includegraphics[width=9cm]{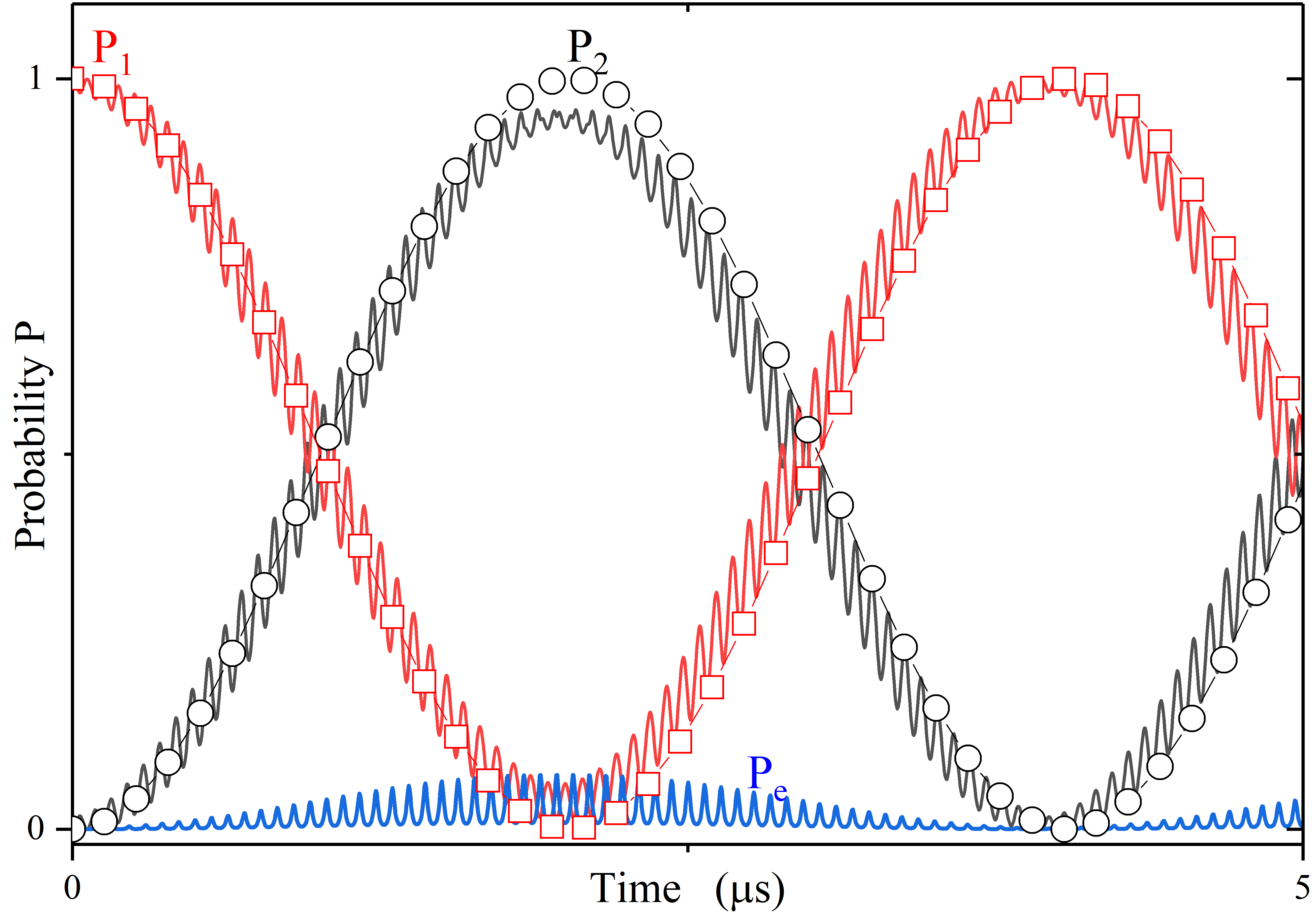}
	\caption{The Rabi oscillations of a single phonon state between two acoustic cavities described by the original Hamiltonian $H_{s}(t)$ (solid oscillating curves) and the Floquet-engineered effective Hamiltonian $H_{12}^{\text{eff}}$. Note that $P_{i}$ ($i=1,2$) represents one phonon being $i$th cavity. The blue oscillating curves denote the probability of the qubit being its excited states. Parameters: $\Delta_{p}/(2\pi)=0.6~\text{GHz}$, $g_{p}/(2\pi)=60~\text{MHz}$, $g_{12}/(2\pi)=1~\text{MHz}$, $\Omega_{d}/(2\pi)=150~\text{MHz}$ and $\lambda=0.5$.}
	\label{fig3m}
\end{figure}

To verify the validation of the above discussions, in the following we numerically compare the time-dependent evolution between the engineered effective Hamiltonian in Eq.~(\ref{Heff}) with the original Hamiltonian 
\begin{equation}
H_{s}(t)=H_{c12}(t)+H_{p}(t).
\end{equation}
In experiments, various methods are employed to enhance the coupling between the SAW modes and superconducting qubits. For example, choosing a strong piezoelectric material will increase the coupling significantly~\cite{Delsing2019}. As discussed in Refs.~\cite{Manenti16,Manenti2017,Moores2018}, we assume $g_{p}/(2\pi)=60~\text{MHz}$. By setting $\Delta_{p}/(2\pi)=0.6~\text{GHz}$, $g_{12}/(2\pi)=1~\text{MHz}$, $\omega_{d}/(2\pi)=15~\text{MHz}$ and $\lambda=0.5$ (corresponding the dashed line position in Fig.~\ref{fig2m}), one finds that $K_{1}\simeq0.25$, and the effective hoping rate $J_{12}$ is about $J_{12}/(2\pi)=0.12~\text{MHz}$. 

We define the states of one phonon being in the $i$th cavity as $|\psi_{i}\rangle=b_{i}^{\dagger}|0\rangle$, where $|0\rangle$ is the vacuum state of the system. 
Employing the above parameters and defining the probabilities $P_{i}=|\langle \Psi_{12}(t)|\psi_{i}(t)\rangle|^{2}$, we plot the evolutions of $P_{1,2}$ in Fig.~\ref{fig3m}, in which the solid curves and symbol curves are governed by the original Hamiltonian $H_{s}(t)$ (solid oscillating curves) and the Floquet-engineered effective Hamiltonian $H_{12}^{\text{eff}}$ (curves with symbols), respectively. We find that, with a period drive on the qubit longitudinal freedom, the two detuned SAW cavities effectively couple together on resonance. Moreover, we plot the probability $P_{e}$ of the qubit being its excited states, and find that $P_{e}\ll1$, indicating that the qubit is hardly excited. Therefore, our Floquet analysis in Eq.~(\ref{H_dis1}) by setting $\sigma_{p}\simeq-1$ is valid. The phase $\phi$ encoded by the longitudinal drive can not be observed in this two-cavity system. In the following, we will demonstrate its effect in phonon-cavity lattice with closed loops, and show how to realize exotic quantum controls of phonons based on the interference effects by breaking time-reversal symmetry of the lattice.

One may wonder whether it is feasible to replace the low-excitation p-qubit (with $\sigma_z\simeq  -1$) by an LC resonator. To discuss this, we first approximately write the superconducting p-qubit Hamiltonian as a resonator with a Kerr nonlinearity as~\cite{Sete2015s,Geerlingsthesis,Campagnethesis}
	\begin{equation}
	H_{q}=\frac{1}{2}\omega_{p}\sigma_{z}^{p}\simeq \omega_{p}c^{\dagger}c+K c^{\dagger}c^{\dagger}cc,
	\end{equation}
	where $c^{\dagger}$ ($c$) is the raising (lowering) operator for the resonator, and $K$ is the resonator anharmonicity. In the limit $K\rightarrow \infty$, we can view $H_{q}$ as an ideal two-level qubit. As discussed in Ref.~\cite{Sete2015s}, for finite $K$, the dispersive coupling between the phonon cavity and the resonator with finite anharmonicity $K$ [Eq.~(\ref{chi_zz})] is modified as
	\begin{equation}
	H_{\text{dis}}=-2\chi'_{0}c^{\dagger}c b_{1}^{\dagger}b_{1}, \qquad \chi'_{0}=-\frac{g_{p}^{2}K}{\Delta_{p}(\Delta_{p}-K)}.
	\label{chi_zzK}
	\end{equation}
	It is found that, in the limit $|K|\gg |\Delta_{p}|$, the dispersive strength $\chi'_{0}$ is equal to $\chi_{0}$ in Eq.~(\ref{chi_zz}). However, in the limit $K=0$, the resonator becomes a linear LC resonator, and the dispersive coupling strength is equal zero, i.e., $\chi'_{0}=0$. Alternatively, one can obtain this by simply considering two coupled resonator in the large detuning regime, which does not result in the cross-Kerr interactions.
	Therefore, by replacing the p-qubit by an LC resonator, we cannot effectively modulate the phonon frequencies and generate the hopping phases in phonon-cavity lattice. Our proposal requires the p-qubit of high anhamonicity with $K\gg \Delta_{p}$.

\section{chiral ground state and phonon circulator}
\begin{figure}[tbph]
	\centering \includegraphics[width=11cm]{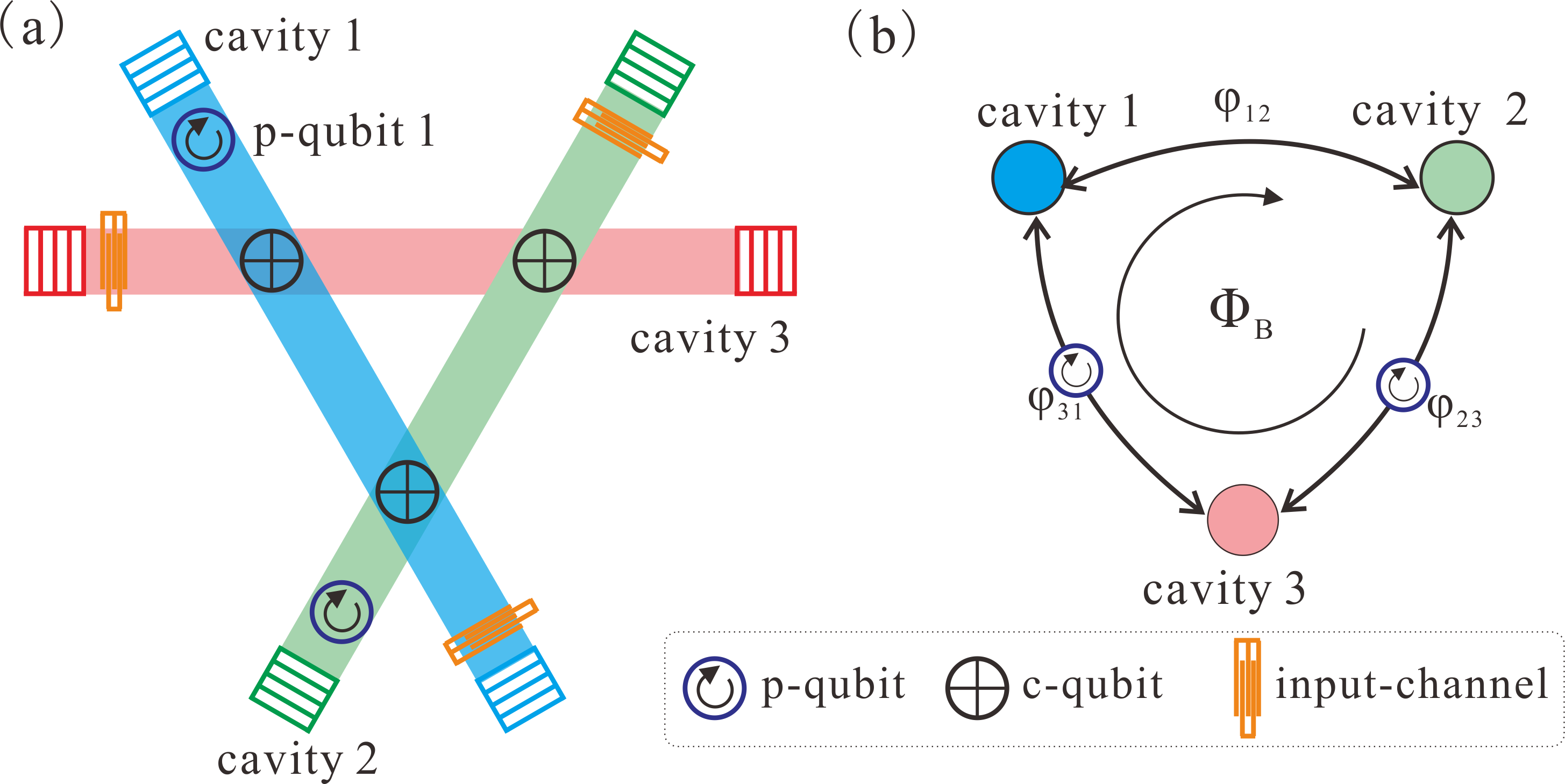}
	\caption{(a) The schematic diagram to realize chiral ground state and phonon circulator in a three-SAW-cavity loop. The couplings between different cavities are mediated by c-qubit. For simplification, we assume cavity 1 (blue) and 2 (green) are on resonant, and they are of the same detuning with cavity 3 (red). The Floquet drivings are applied though two p-qubits (located in cavity 1 and 2, respectively), where the hopping phase $\phi_{13}$ ($\phi_{23}$) between cavity 3 and 1 (2) is generated via the methods described in Sec. III. (b) The generated synthetic magnetism $\Phi_{B}=\phi_{12}+\phi_{23}+\phi_{31}$ ($\phi_{12}=0$) in the three-SAW-cavity loop via Floquet engineering.}
	\label{fig4m}
\end{figure}

To observe the effects of breaking TRS in a coupled SAW cavity systems, we first consider the simplest three-site system forming as a closed loop depicted in Fig.~\ref{fig4m}: the SAW cavities are mediated via the c-qubits. For simplification, we assume that cavity 1 and 2 are resonantly coupled ($\delta_{12}=0$), while of the same detuning with cavity 3, i.e., $\delta'_{13}=\delta'_{23}$. Moreover, two p-qubits are placed into cavity 1 and 2 separately, which are employed to induce the required cavity frequency modulations. The phases $\phi_{13}$ ($\phi_{23}$) is encoded into the longitude drives of p-qubit 1 (2). Under the rotating frame with the unitary transformation operator $$U_{3R}(t)=\exp{\left[-i(\omega_{m3}\sum_{i=1,2,3}b_{i}^{\dagger}b_{i})t\right]}\exp{\left[-i(\omega_{m3}\sum_{i=1,2}\sigma_{pi}^{z})t\right]},$$ the total Hamiltonian in the interaction picture is written as:
\begin{gather}
H_{3R}=H_{0}+H_{pm}+H_{c}; \label{Htot3}\\
H_{0}=\frac{1}{2}\sum_{i=1,2}\left[(-\Delta_{pi}+\delta'_{i3})\sigma_{pi}^{z}+\delta'_{i3}b_{i}^{\dagger}b_{i}\right];\\
H_{pm}=\sum_{i=1,2}\left[g_{pi}(b_{i}^{\dagger}\sigma_{pi}^{-}+b_{i}\sigma_{pi}^{+})-\Omega_{pi}\sigma_{pi}^{z}\cos(\omega_{d}t+\phi_{i})\right]; \\
H_{c}=[(g_{12}b_{1}^{\dagger}b_{2}+g_{23}b_{2}^{\dagger}b_{3}+g_{13}b_{1}^{\dagger}b_{3})+\text{H.c.}]
\end{gather}
where $\Delta_{pi}$ ($g_{pi}$) is the detuning (coupling strength) between the $i$th phonon cavity and $i$th p-qubit, $\delta'_{ij}$ ($g_{ij}$) is the detuning (coupling strength) between cavity $i$ and $j$, and $\Omega_{pi}$ ($\phi_{i}$) is the driving strength (phase) on the $i$th p-qubit.

As discusses in Sec.~III, the qubit longitudinal drive will modulate the cavity frequencies periodically, and it is reasonable to consider only first-order frequency component. Consequently, we can eliminate the degree of the freedom for the p-qubits, and the system Hamiltonian is written as 
\begin{eqnarray}
H_{3R}(t)&=&[(g_{12}b_{1}^{\dagger}b_{2}+g_{23}b_{2}^{\dagger}b_{3}e^{i\delta_{23}t}+g_{13}b_{1}^{\dagger}b_{3}e^{i\delta_{13}t}
) \notag \\
&&+\text{H.c.}]+\sum_{i=1,2}K_{i}\omega_{d}\cos(\omega_{d}t+\phi_{i})b_{i}^{\dagger}b_{i},
\label{HSS}
\end{eqnarray}
where $\delta_{13}$ and $\delta_{23}$ are shifted frequency, which can be calculated according to the relation displayed in Eq.~(\ref{detune12}). As shown in Eq.~(\ref{knn}), $K_{i}$ is expressed as 
\begin{equation}
K_{i}=\frac{g_{pi}^{2}}{\Delta_{pi}\omega_{d}}\xi_{1i},
\end{equation}
with $\xi_{1i}$ being the first-order amplitude of the periodic function $1/[1+(\Omega_{pi}/\Delta_{pi})\cos(\omega_{d}t+\phi_{i})]$.

By setting $\delta_{13}=\delta_{23}=-\omega_d$, we obtain	
\begin{eqnarray}
H_{3R}(t)&=&g_{12}(b_{1}^{\dagger}b_{2}+\text{H.c.})+
\sum_{i=1,2} \left[g_{i3}(b_{i}^{\dagger}b_{3}e^{-i\omega_{d}t}+\text{H.c.})+K_{i}\omega_{d}\cos(\omega_{d}t+\phi_{i})b_{i}^{\dagger}b_{i}\right].
\label{H3RR}
\end{eqnarray}	
Similar to the discussions from Eq.~(\ref{hrot}) to Eq.~(\ref{Hefflong}), we perform the unitary transformation to $H_{3R}(t)$, 
\begin{eqnarray}
U(t)=R_{b1}(t)R_{b2}(t), \quad R_{bi}(t)=\sum_{k=-\infty}^{k=\infty}\mathcal{J}_{k}(K_{i}b_{i}^{\dagger}b_{i})e^{ik(\omega_{d}t+\phi_{i})}.
\end{eqnarray}
The transformed Hamiltonian is expressed as 
\begin{eqnarray}
H'_{3R}(t)&=&g_{12}\left[R_{b1}(t)b_{1}^{\dagger}R_{b1}^{\dagger}(t)R_{b2}(t)b_{2}R_{b2}^{\dagger}(t)+\text{H.c.}\right] \notag\\
&&+
\sum_{i=1,2} g_{i3}\left[R_{bi}(t)b_{i}^{\dagger}R_{bi}^{\dagger}(t)b_{3}e^{-i\omega_{d}t}+\text{H.c.}\right],
\label{H3RU00}
\end{eqnarray}
Since phonon cavity 1 and 2 are resonantly coupled, we can approximately write $R_{b1}(t)=R_{b2}(t)\simeq 1$ for the first term in Eq.~(\ref{H3RU00}). However, due the cavity 1 (2) and 3 are of large detuning, we should expand $R_{bi}(t)$ to the higher-order for the second term. Similar to discussions from Eq.~(\ref{Heff}) to Eq.~(\ref{Hefflong}),
we obtain
\begin{eqnarray}
H_{3R}(t)=
g_{12}\left[b_{1}^{\dagger}b_{2}+\text{H.c.}\right]
+\sum_{i=1,2} g_{i3}\frac{K_{i}}{2}\left[b_{i}^{\dagger}b_{3}e^{\phi_{i}}+\text{H.c.}\right]+g_{i3}\left[b_{i}^{\dagger}b_{3}e^{-i\omega_{d}t}+\text{H.c.}\right],
\label{H3RU2}
\end{eqnarray}
where the last term indicates that there is another hopping channel between cavity 1 and 2 mediated by their detuning couplings with cavity 3. Similar with discussions in Sec.~II, we employ the effective Hamiltonian methods in Ref.~\cite{James2007}, and obtain the corresponding hoping rate as $(-g_{13}g_{23})/\omega_{d}$. We write the Hamiltonian~(\ref{H3RU2})
in a symmetric form
\begin{gather}
H_{3R}^{\text{eff}}=(J_{12}e^{i\phi_{12}}b_{1}^{\dagger}b_{2}+J_{23}e^{i\phi_{23}}b_{2}^{\dagger}b_{3}+J_{31}e^{i\phi_{31}}b_{3}^{\dagger}b_{1})+\text{H.c.} \label{Heff1} \\
\quad J_{12}=g_{12}-\frac{g_{13}g_{23}}{\omega_{d}},  \qquad J_{23}=\frac{g_{23}K_{2}}{2}, \quad J_{31}=\frac{g_{13}K_{1}}{2}, \\
\phi_{12}=0, \quad \phi_{23}=\phi_{2}, \quad \phi_{31}=-\phi_{1}.
\end{gather}

Due to the resonant hopping between cavity 1 and 2, their coupling phase is fixed as $\phi_{12} = 0$ in our proposal. Note that the resonant coupling strength $J_{12}$ between cavity 1 and 2 also needs to be shifted due to coupling with cavity 3.

\begin{figure*}[tbph]
	\centering \includegraphics[width=17cm]{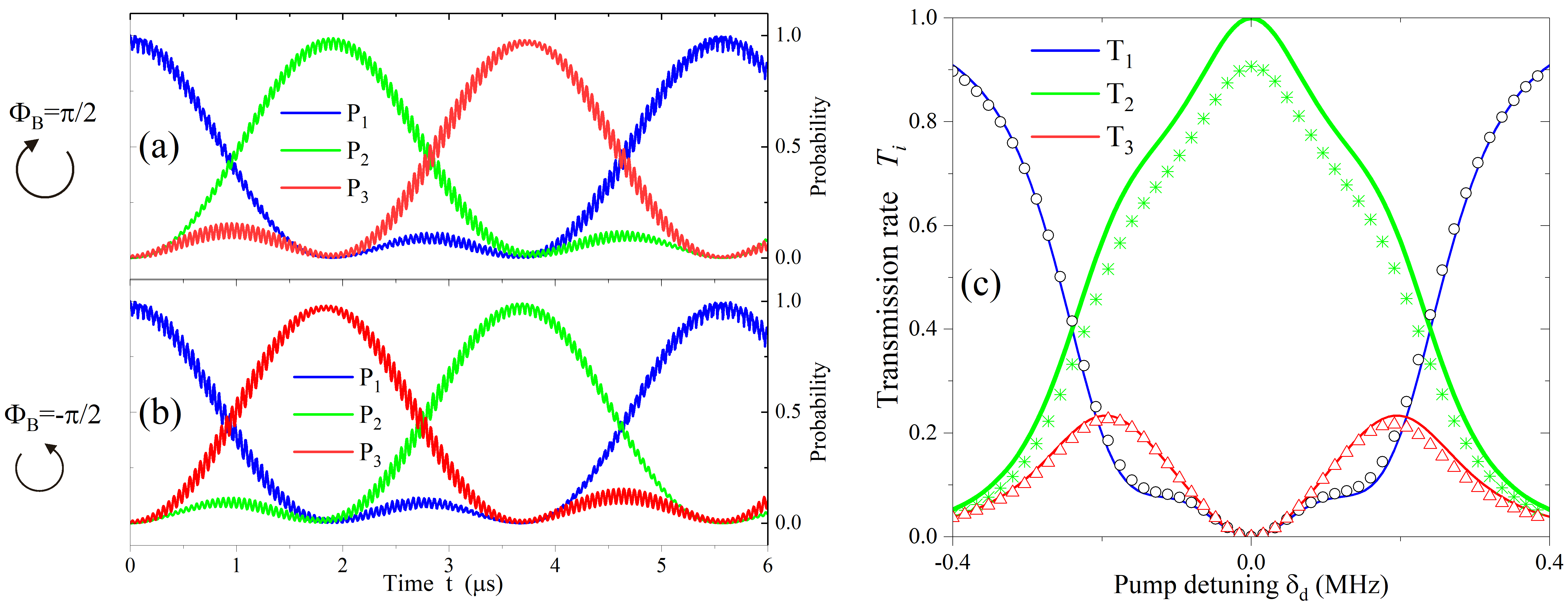}
	\caption{(a) Time evolutions for single-phonon states in three SAW cavities. Since TRS is broken, for $\Phi_{B}=\pi/2$, the single phonon will circulate along the counter-clockwise direction in the sequence $\text{...1 (blue)} \rightarrow \text{2 (green)}\rightarrow \text{3 (red)}\rightarrow \text{1 (blue)...}$. (b) corresponds to the opposite case with  $\Phi_{B}=-\pi/2$, and the single phonon will circulate in the opposite direction to (a). (c) Phonon circulator behavior changes with pump frequency detuning $\delta_{d}$. We assume that the input the acoustic signal from the IDT in cavity 1 (blue curve) for $\Phi_{B}=\pi/2$. Around $\delta_{d}=0$, most of output signal power is routed output channel of cavity 2 (green) along the counter-clockwise direction. All the curves in (a, b) and the dashed curves in (c) are plotted according to the Floquet Hamiltonian $H_{3R0}$ in Eq.~(\ref{Htot3}), rather than the effective coupling Hamiltonian $H_{3R}^{\text{eff}}$ in Eq.~(\ref{Heff1}). The solid curves in (c) is plotted according to analytical results in Eq.~(\ref{anamat3R}). The adopted parameters are: $\Delta_{p1}=\Delta_{p2}=600~\text{MHz}$, $g_{p1}=g_{p2}=60~\text{MHz}$,
			$\Omega_{p1}=\Omega_{p2}=150~\text{MHz}$,
			$\omega_{d}=-20~\text{MHz}$, $g_{13}=g_{23}=1.1~\text{MHz}$, 
			$g_{12}=0.042~\text{MHz}$ and $\kappa=0.2~\text{MHz}$.} 
	\label{fig5m}
\end{figure*}

Similar to a charged particle moving on a lattice, the phonon can also accumulate analogous Peierls phase $\phi_{ij}=\int_{i}^{j}\mathbf A d\mathbf r$~\cite{Luttinger1951}, where $\mathbf A$ is the synthetic vector potential~\cite{Ozawa2019,Schmidt2015}. Therefore, in a closed loop $C$, the sum $\sum_{C}\phi_{ij}$ is equal to the synthetic magnetic flux $\Phi_{B}$ through the loop. In our proposal, the synthetic magnetic flux can be conveniently tuned by the phase of  p-qubit longitudinal drives.

Once the phase information is encoded into the closed loop of a SAW lattice, the dynamics of the whole system will change dramatically~\cite{Roushan2016}. We first discuss the situation of breaking TRS of the Hamiltonian. Since the phonon number states should be time-reversal invariant, we require the annihilation operator $b_{i}$ satisfies the following equation~\cite{Koch2010}
\begin{equation}
\Theta b_{i}\Theta^{-1}=e^{i\vartheta_{i} }b_{i},
\end{equation}
where $\Theta$ is the time-reversal operator, which is antilinear and antiunitary. After performing time-reversal transformation, the operator $b_{i}$ acquires a phase $\vartheta_{i}$.
Given that 
\begin{equation}
\Theta H_{3R}^{\text{eff}}\Theta^{-1}=H_{3R}^{\text{eff}}
\end{equation}
is valid, the effective Hamiltonian is time-reversal invariant, which requires the sum of the coupling phase along a closed path satisfies~\cite{Koch2010}
\begin{equation}
\Phi_{B}=\sum_{C}\phi_{ij}=\mathbb{Z}\pi.
\label{chi_cond}
\end{equation}
Otherwise, the TRS of the system will be broken, and one may observe acoustic chiral transport. For simplification, we set the hopping rates identically as $J_{ij}=J$. It can be proved that, for arbitrary $\phi_{ij}$, the following gauge transformation~\cite{Koch2010} 
\begin{equation}
\Theta b_{i}\Theta^{-1}=e^{i\vartheta_{i} }b_{i}, \quad \vartheta_{i}=(1)^{i}(\frac{\Phi_{B}}{3}-\phi_{3i}), \quad i=1,2
\end{equation}
will transfer the Hamiltonian in Eq.~(\ref{Heff1}) with one phase factor $\phi_{c}=\Phi_{B}/3$, i.e.,
\begin{equation}
H_{3R}^{\text{eff}}=Je^{\phi_{c}}(b_{1}^{\dagger}b_{2}+b_{2}^{\dagger}b_{3}+b_{3}^{\dagger}b_{1}+\text{H.c.}).
\end{equation}
where we assume $J_{12}=J_{23}=J_{31}=J$.
Therefore, only $\Phi_{B}$ can produce observable physical effects.
In Fig.~\ref{fig5m}(a, b), by assuming $\phi_{c}=\pm\pi/6$ ( i.e., $\Phi_{B}=\pm\pi/2$), we plot the probabilities $P_{i}$ of the single photon being in the $i$th cavity changing with time governed by the orginal Floquet Hamiltonian in Eq.~(\ref{Htot3}), i.e., by considering the degree of freedom of the p-qubits.
 The initial state is assumed to be one phonon in cavity 1. In Fig.~\ref{fig5m}(a), we clearly find that, for $\Phi_{B}=\pi/2$, the single phonon will propagate in the circular order 
$$\text{...1 (blue)} \rightarrow \text{2 (green)}\rightarrow \text{3 (red)}\rightarrow \text{1 (blue)...} $$
Therefore, the acoustic energy will propagate along the counter-clockwise direction. The clockwise direction is forbidden for $\Phi_{B}=\pi/2$, but allowed for the opposite flux $\Phi_{B}=-\pi/2$, which is shown in Fig.~\ref{fig5m}(b). Therefore, The chiral transfer of the single phonon indicates that the synthetic flux $\Phi_{B}$ can be employed for realizing phonon circulator.

Analog to a photon circulator~\cite{Koch2010}, a phonon circulator is an acoustic element with three or more ports~\cite{Habraken2012,Shen2018}, and will transport clockwise (or counterclockwise) the input in the $j$th port as the output in the $(j\pm1)$th port. As shown in Fig.~\ref{fig5m}(a), for SAW experimental implementations, the acoustic in/out- port are usually made of IDTs, which can convert the electromagnetic signals into the acoustic signals, and vice versa.
We assume the input field operator for the $j$th cavity is $c_{\text{in},j}$. In the rotating frame of the input frequency, the interaction between the input field and the system reads~\cite{Collett84,Gardiner85}
\begin{equation}
H_{\text{in}}=i\sqrt{\kappa}\sum_{j=1}^{3}(c_{\text{in},j}^{\dagger}b_{j}-c_{\text{in},j}b_{j}^{\dagger})
\end{equation}
where $\kappa$ is the phonon escaping rate into the input/output channels. Defining the detuning between the input pump field and the cavity 1 as $\delta_{d}$, the Heisenberg equation of the intracavity operator is
\begin{eqnarray}
\dot{\mathbf B}(t)&=&-iJ \mathbf \Gamma \mathbf B (t)-(i\delta_{d}+\frac{\kappa}{2})\mathbf B (t)-\sqrt{\kappa}  \mathbf C_{\text{in}},
\end{eqnarray}
where $\mathbf B=[b_{1}, b_{2}, b_{3}]^{T}$ and $\mathbf C_{\text{in}}=[c_{\text{in},1}, c_{\text{in},2}, c_{\text{in},3}]^{T}$ are the intracavity and input field operator vectors, and $\mathbf \Gamma$ is the coupling matrix:
\begin{eqnarray}
\mathbf \Gamma&=&\left[
\begin{matrix}
0 & e^{i\phi_{c}} & e^{-i\phi_{c}}\\
e^{-i\phi_{c}} & 0  & e^{i\phi_{c}}\\
e^{i\phi_{c}} & e^{-i\phi_{c}} & 0 
\end{matrix}
\right],
\end{eqnarray}
The above linear equation can be easily solved in the frequency domain, and the solution reads
\begin{equation}
\mathbf B(\omega)=\frac{-\sqrt{\kappa}}{i(\omega+\delta_{d})+iJ\mathbf \Gamma+\frac{\kappa}{2}}\mathbf C_{\text{in}}.
\end{equation}
The output can be obtained via the input-output relation $c_{\text{out},i}=\sqrt{\kappa}b_{i}(t)+c_{\text{in},i}$~\cite{Collett84,Gardiner85}. Specifically, we pay attention to the transmission relation when the system reaches its steady state with $t\rightarrow\infty$ (corresponding to $\omega=0$). Under these conditions, the output field is written as
\begin{equation}
\mathbf C_{\text{out}}=\mathbf S(\delta_{d}) \mathbf C_{\text{in}}, \quad \mathbf S(\delta_{d}=0)=\frac{i\delta_{d}+iJ\mathbf \Gamma-\frac{\kappa}{2}}{i\delta_{d}+iJ\mathbf \Gamma+\frac{\kappa}{2}},
\label{anamat3R}
\end{equation}
where $\mathbf S(\delta_{d})$ is the scattering matrix for the phonon circulator. For an ideal 
circulator, we require 
\begin{eqnarray}
|\mathbf S(\delta_{d}=0)&|=&\left[
\begin{matrix}
0 & 0 & 1\\
1& 0  & 0\\
0 & 1 & 0 
\end{matrix}
\right],
\label{smatrix}
\end{eqnarray}
which describes the phonon flux injected into a port will be transferred into next one circularly~\cite{Fleury2014,Gu2017}. It can be proved that, only under the conditions $\phi_{c}=\pi/6$ and $J=\kappa/2$, the scattering matrix is equal to the ideal circulator matrix in Eq.~(\ref{smatrix}). We assume that only cavity 1 is injected with an input field, i.e., $\mathbf C_{\text{in}}=[c_{\text{in},1}, 0, 0]^{T}$. In Fig.~\ref{fig5m}(c), we plot the transmission rate $T_{i}=|c_{\text{out,i}}/c_{\text{in},1}|^{2}$ for $i=1,2,3$ (corresponding to curves with symbols) governed by original Hamiltonian~(\ref{Htot3}), which match well with the analytical results [the solid curves, derived from analytical results in Eq.~(\ref{anamat3R})].
It can be found that, at $\delta_{d}=0$, the injected phonon flux of cavity 1 is effectively circulated into the output port 2, with little power leaking into the other two channels. Therefore, by driving the coupled p-qubits, one can effectively realize a phonon circulator by generating synthetic magnetic flux in an acoustic system.

\section{Simulating acoustic Aharonov-Bohm effects}
\begin{figure}[tbph]
	\centering \includegraphics[width=12cm]{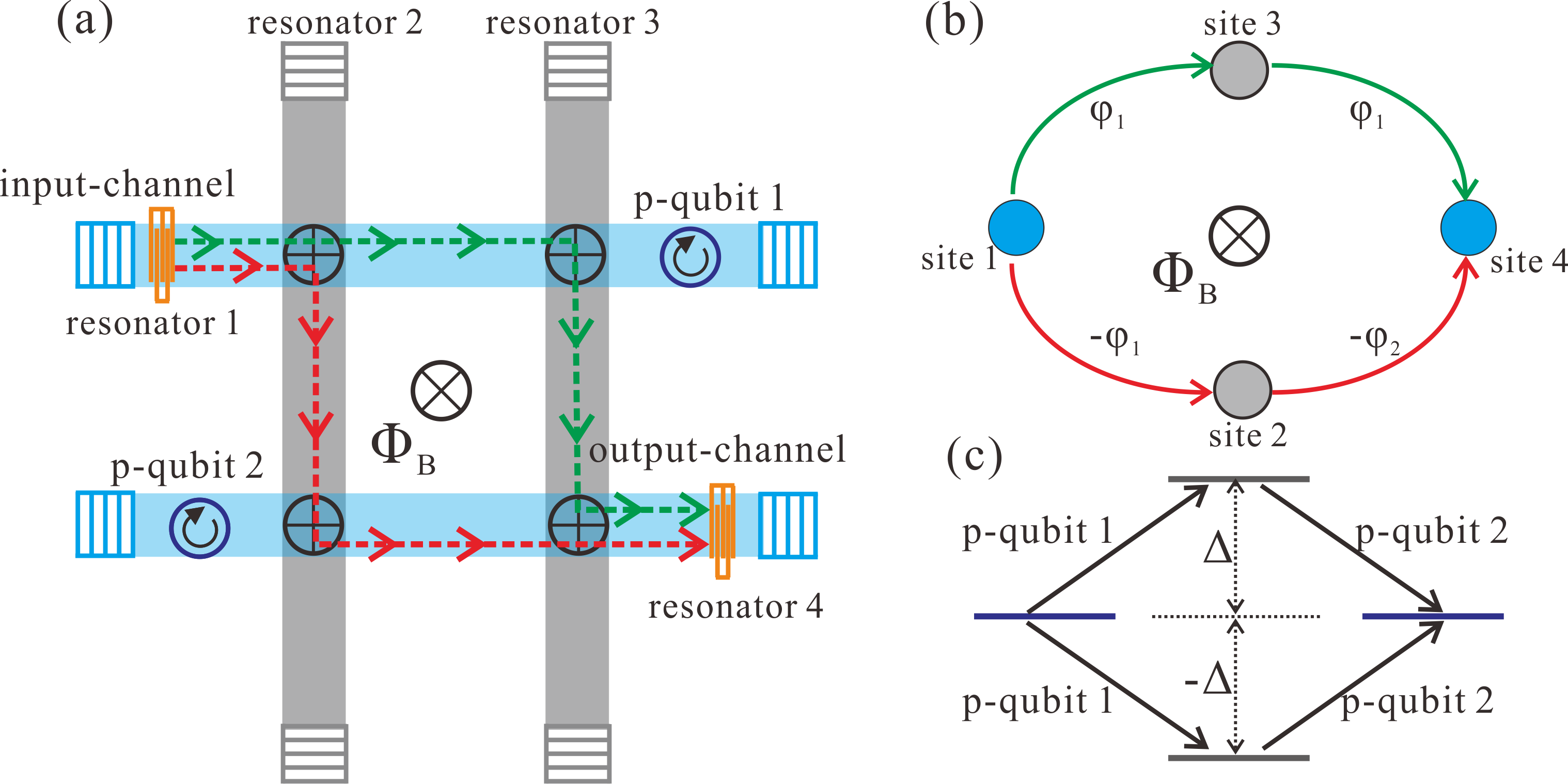}
	\caption{(a) The schematic diagram to observe the acoustic Aharonov-Bohm effects in a four-SAW-cavity acoustic loop: the acoustic wave transport from site 1 to 2 (blue),  are composed by two ``paths", which are through cavity 2 or 3, respectively. The IDT in cavity 1 and 4 are the acoustic source and detector, respectively. The propagating phonons will feel the existence of the synthetic flux $\Phi_{B}$. (b) The synthetic magnetism in the acoustic loop, $\Phi_{B}=2(\phi_{1}+\phi_{2})$, are generated by Floquet engineering the p-qubit in cavity 1 and 4 simultaneously. (c) The energy detuning relations between the four cavities: cavity 1 and 4 are on resonance, while 2 and 3 are detuned with them of amount $\Delta_{AB}$ and $-\Delta_{AB}$, respectively.}
	\label{fig6m}
\end{figure}
In the experimental demonstration of the AB effect, an electron beam is split into two paths (denoted as $C_{1,2}$). According to the interaction Hamiltonian describing electron moving in an electromagnetic field, in addition to the standard propagation phase, the electron in two paths will acquire phases for different paths due to the magnetic field:
\begin{equation}
\widetilde{\phi_{i}}=-\frac{e}{\hbar}\int_{C_{i}} \mathbf {A}d \mathbf{l}, \quad i=1,2.
\end{equation}

It should be noted that $\Delta\widetilde{\phi_{i}}$ has a gauge-dependent form. However, for a closed loop, the phase difference 
\begin{equation}
\Delta\widetilde{\phi}=\widetilde{\phi_{1}}-\widetilde{\phi_{2}}=\Phi_{B}
\end{equation}
is gauge invariant, and always equals to the magnetic flux through the loop, which can produce observable effects. 

Since acoustic phonons are neutral particles with $e=0$, they cannot feel the existence of the magnetic field. However, as discussed in previous sections, one can realize artificial magnetism via the  qubit-assisted Floquet engineering. To simulate the analogous AB effects for acoustic waves, we propose a simple four-site lattice model in Fig.~\ref{fig6m}: two resonant acoustic cavities 1 and 4 (blue ones), are hopped with cavity 2 and 3 and forming a closed loop. 
Resonators (2 and 3) are detuned with 1 and 4 with freuqency $\Delta_{AB}$ and $-\Delta_{AB}$, respectively [as depicted in Fig.~\ref{fig6m}(c)]. 
\begin{figure}[tbph]
	\centering \includegraphics[width=16cm]{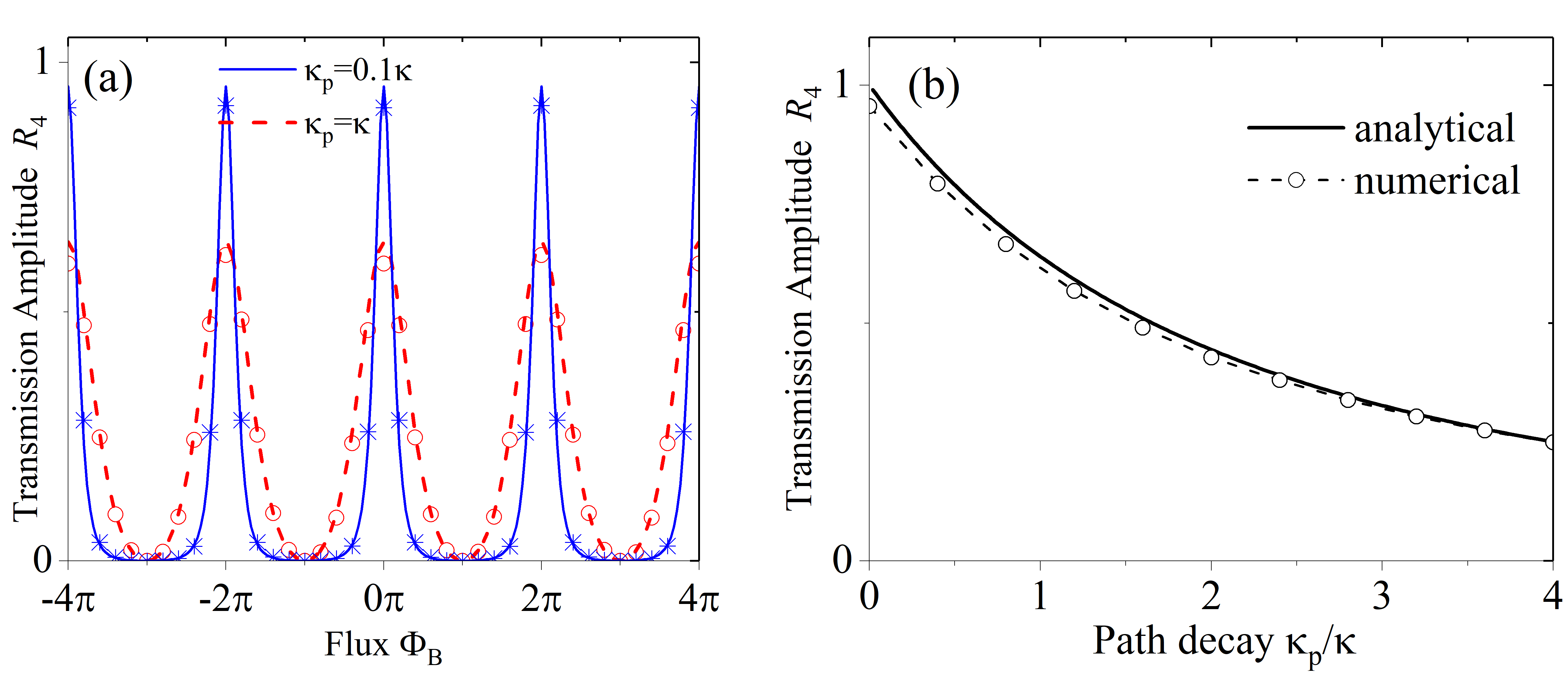}
	\caption{(a) The interference patten of the acoustic Aharonov-Bohm (AB) effects by changing the synthetic magnetism $\Phi_{B}$ in Fig.~\ref{fig6m}. The blue (red) curve adopts ``path" decay rate as $\kappa_{p}=0.1\kappa$ ($\kappa_{p}=\kappa$), respectively. The symbol (solid) curve corresponds to the numerical (analytical) results governed by Floquet Hamiltonian in Eq.~(\ref{FABe}) [analytical expression in Eq.~(\ref{anamat})]. (b) The amplitudes of the constructive interference peaks in acoustic AB effects changes with the ``path" decay rate $\kappa_{p}$. The adopted parameters are:
		$\Delta_{AB}=-20~\text{MHz}$, $g_{ij}=g=1~\text{MHz}$, $K_{1}=K_{2}=0.2$, and $\kappa_{1,2}=\kappa=0.2~\text{MHz}$.}
	\label{fig7m}
\end{figure}

In this case, cavity 1 and 4 are parametrically driven by the p-qubits at frequency $\Delta_{AB}$ with relative phases $\widetilde{\phi}_{1}$ and $\widetilde{\phi}_{4}$, respectively. The Floquet Hamiltonian is written as 
\begin{eqnarray}
H_{AB}(t)&=&\sum_{i=2,3}\sum_{j=1,4}g_{ij}(b_{i}^{\dagger}b_{j}e^{i\delta_{ij}t}+\text{H.c.}) \notag \\
&&+\sum_{j=1,4}K_{j}\Delta_{AB}\cos(\Delta_{AB}t+\phi_{j})b_{j}^{\dagger}b_{j},
\label{FABe}
\end{eqnarray}
where $\delta_{31}=-\delta_{21}=\Delta_{AB}$ and $\delta_{34}=-\delta_{24}=\Delta_{AB}$. 
As discussed in Sec. II, due to the opposite detuning relations, the hopping phases between 1 (4) and 2, are of the opposite signs with 1(4) and 3, which is as shown in Fig.~\ref{fig6m}(b). Assuming $g_{ij}=g$, $K_{i}=K$ and $J\simeq gK/2$, the effective Hamiltonian reads
\begin{equation}
H_{AB}^{\text{eff}}=\sum_{j=1,4}J[b_{j}(b_{2}^{\dagger}e^{i\phi_{j}}+b_{3}^{\dagger}e^{-i\phi_{j}})+\text{H.c.}].
\end{equation}
Therefore, there will be two paths for a phonon transferring from cavity 1 to 4, i.e., either $1\rightarrow3\rightarrow4$ or $1\rightarrow2\rightarrow4$. The whole hybrid system in Fig.~\ref{fig6m}(a) can be viewed as a discrete proposal for simulating phonon AB effect.  Analog to an electron moving in the continuous space, the synthetic magnetism $\Phi_{B}=2(\widetilde{\phi}_{1}+\widetilde{\phi}_{4})$ affecting on acoustic waves (marked with green and red arrows in Fig.~\ref{fig6m}, respectively) will also produce observable interference effects. The decoherences $\kappa_{2,3}=\kappa_{p}$ of cavity 2 and 3, can be viewed as the noise in the propagating paths, which will keep on ``kicking" on the itinerant phonons and leading to the dephasing of the phonons.

We assume that the cavity 1 is resonantly driven via a coherent input field $c_{\text{in},1}$. We set the hopping between each site of the same strength $J$. Similar with the discussions of a phonon circulator in Sec. III, the transmission rate can be solved via the following equation in the frequency domain
\begin{equation}
\mathbf C_{\text{out}}=\frac{2iJ\mathbf \Gamma_{\text{AB}}-\mathbf K}{2iJ\mathbf \Gamma_{\text{AB}}+\mathbf K} \mathbf C_{\text{in}}, 
\label{anamat}
\end{equation}
where $\mathbf C_{\text{out(in)}}$ is the output (input) field operator vector of the four sites, and the coupling and decoherence matrices are written as 
\begin{gather}
\mathbf \Gamma_{\text{AB}}=\left[
\begin{matrix}
0 & e^{i\widetilde{\phi}_{1}} & e^{-i\widetilde{\phi}_{1}} & 0\\
e^{-i\widetilde{\phi}_{1}} & 0 & 0  & e^{i\widetilde{\phi}_{4}}\\
e^{i\widetilde{\phi}_{1}}  & 0  & 0 & e^{-i\widetilde{\phi}_{4}} \\  
0  & e^{-i\widetilde{\phi}_{4}}  & e^{i\widetilde{\phi}_{4}} & 0 
\end{matrix}
\right], \\
\mathbf K =\rm{diag}\left[\kappa_{1} ,\kappa_{2} ,\kappa_{3} ,\kappa_{4} \right].
\label{fourmatrix}
\end{gather}
 
The interference patterns versus the flux $\Phi_{B}$ for different path loss $\kappa_{p}$ are shown in Fig.~\ref{fig7m}(a), where we have assumed $\kappa_{1,2}=\kappa$. One can find that,
when $\Phi_{B}=n\pi$ (n is an odd integer), the two transmission paths are destructive, and there is no phonon collected by the output port of cavity 4. At $\Phi_{B}/\pi=2n\pi$, the interference becomes constructive, resulting in the largest phonon output. Moreover, as shown in Fig.~\ref{fig7m}(b), the highest transmission rate will decrease due to the increase of the path loss $\kappa_{p}$, and the width of the interference patten will also be enlarged. In such a way, the electron AB effect can be effectively reproduced in an acoustic system.

\section{outlooks and discussions}
\begin{figure}[tbph]
	\centering \includegraphics[width=9cm]{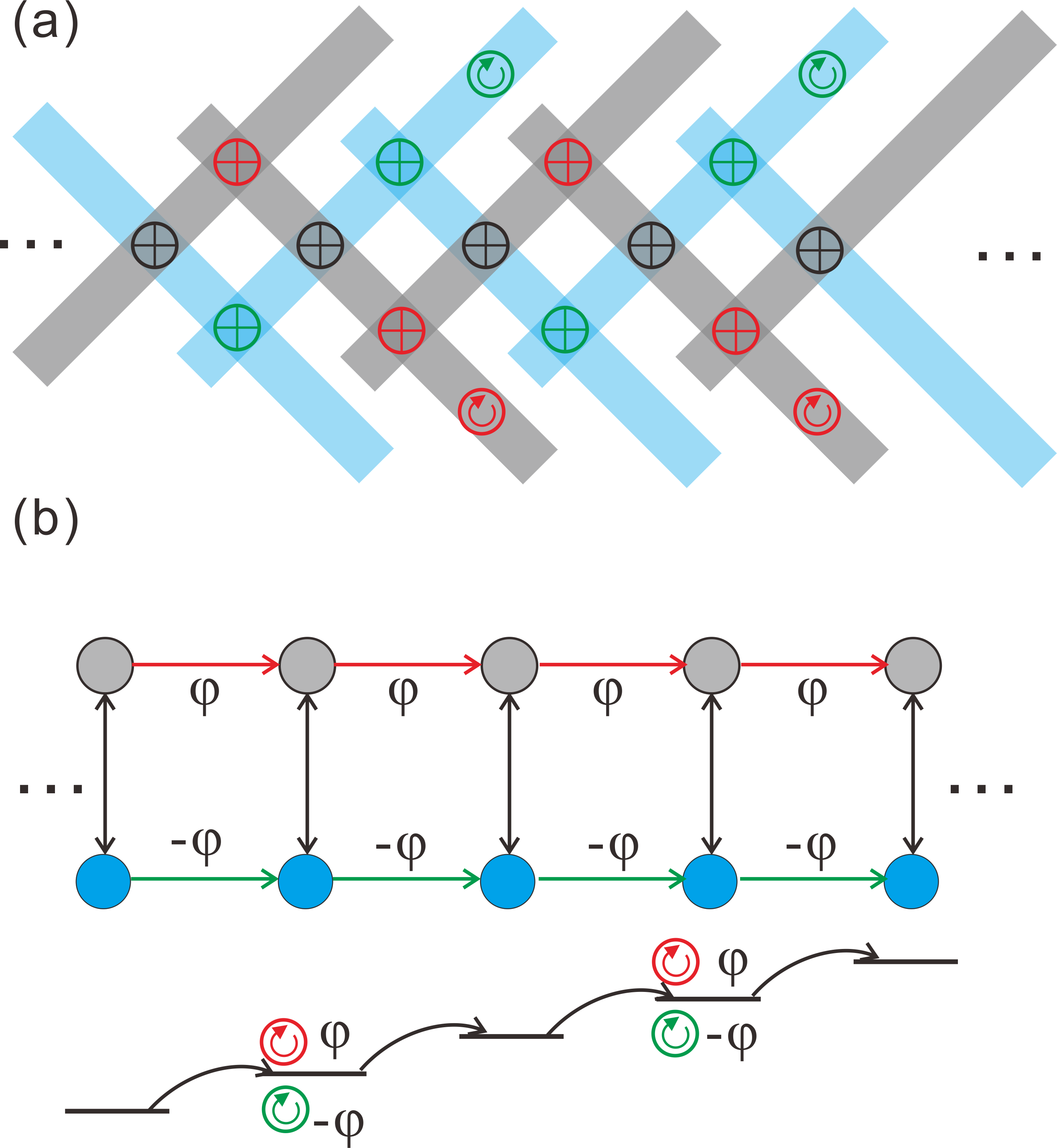}
	\caption{(a) The schematic diagram to simulate Hofstadter ladder in a SAW-cavity-based lattice. The acoustic gauge field is generated via the Floquet driving between the neighbor sites in each leg. the p-qubit is placed in the even-numbered SAW-cavity. The chain of grey (blue) cavity plays the role of the upper (lower) leg of the Hofstadter ladder. (b) The Hofstadter ladder model for (a). The red (green) p-qubit in $j$-th cavity for the upper (lower) leg will induce a hopping phase $\phi$ ($-\phi$), which results in an acoustic synthetic magnetism $\Phi_{B}=2\phi$ in each square lattice.}
	\label{fig8m}
\end{figure}
In the present of a magnetic field, the two-dimensional electron gas has equally spaced landau level. The discrete lattice version of the landau-level is the Hofstadter model~\cite{Hofstadter1976,Ozawa2019}. Assuming the he Landau gauge along the
y direction and the vector potential as $\mathbf {A}=(0,x\phi,0)$, the Hamiltonian of this tight-binding model with a squire lattice is 
\begin{equation}
H_{\text{Hof}}=\sum_{x,y}(-te^{iy\phi}a_{x+1,y}^{\dagger}a_{x,y}+t'a_{x,y}^{\dagger}a_{x,y+1}+\text{H.c.}),
\label{Hhof}
\end{equation}
where $t$ ($t'$) is the hopping strength along (vertical to) the Landau gauge direction. The interaction Hamiltonian $H_{\text{hof}}$ describes a non-zero magnetic flux per plaquette of the lattice. The paths interference of charged particles will lead to a fractal energy spectrum as a function of $\phi$, which is well known as the Hofstadter butterfly~\cite{Thouless1982}. This model exhibits a rich structure of topological invariant depending on $\phi$. Due to this, the quantum simulation of the Hofstadter model has been extensively studied both theoretically and experimentally~\cite{Kuhl1998,Jaksch2003,Owens2018}. 

According to Refs.~\cite{Hgel2014,Tai2017}, one ladder of the Hofstadter model [i.e., by setting $y=1,2$ in Eq.~(\ref{Hhof})] can realize effective spin-orbit coupling, and can approximately reproduce the energy spectrum of the chiral edge states of the two-dimensional model. 
In Fig.~\ref{fig8m}, based on previous discussions, we present a proposal to realize one ladder of the Hofstadter model in phonon-based lattice: The gray ($a_{i}$) and blue ($b_{i}$) acoustic cavities in Fig.~\ref{fig8m}(a) correspond to the upper and lower legs in Fig.~\ref{fig8m}(b), respectively. The coupling between $a_{i}$ and $b_{i}$ are assumed resonantly mediated by the c-qubits with strength $t$.
As shown in Fig.~\ref{fig8m}(b), the detuning between site $i$ and $i+1$ is of the same amount, which is similar to Wannier-Stark ladders in the quantum simulations based on ultracold atoms in optical lattices~\cite{Aidelsburger2013,Miyake2013,Goldman2015}.

In the upper (lower) leg, the p-qubit is assumed to be placed in the even site $j$ for Floquet engineering with phase $\phi^{+}$ ( $\phi^{-}$), where we have assumed each p-qubit is encoded with the same phase. For simplification, we set $\phi^{+}=-\phi^{-}=\phi$.
As discussed in Sec. II, the hopping from site $j-1\rightarrow j$ and $j\rightarrow j+1$ will be encoded with the same phase. We can write the Hamiltonian for the ladders as 
\begin{equation}
H_{\text{lad}}=\sum_{i}(t'e^{i\phi}a_{i+1}^{\dagger}a_{i}+t'e^{-i\phi}b_{i+1}^{\dagger}b_{i}+Ja_{i}^{\dagger}b_{i})+\text{H.c.},
\label{Hoff}
\end{equation}
where we have assumed the coupling strength $t'$ in each ladder leg are the same. 
As discussed in Ref.~\cite{Hgel2014}, employing the Hamiltonian in Eq.~(\ref{Hoff}), we can simulate the chiral edge states two-dimensional Hofstadter model in a phonon-based lattice system. 

\section{Conclusions}
We propose a novel method to generate synthetic magnetisms for phonons via quantum control a coupled qubit. We find that, by applying a longitudinal driving on the qubit, the frequency of the mechanical mode will be periodically modulated. The Floquet engineering on the qubit can tailor the detuned phonon-phonon coupling as a resonant interaction.
By analyzing the frequency components of the modulation, we find that the phase encoded into the longitudinal drive can be exactly transferred into the two-cavity hopping, which allows to create synthetic magnetism for phonons in nanoscale platforms.
In our discussions, we take the SAW cavities for the example. However, this proposed mechanism is general, and can be applied to various types of nano- or micro-scale hybrid quantum systems involving photons and even other neutral polaritons. 

The time-reversal symmetry in phonon lattice can be conveniently controlled by the generated synthetic magnetism flux. Based on this mechanism, we propose a method to realize the phonon circulators in acoustic systems.
Moreover, we show how to observe the acoustic Aharonov-Bohm effects.
In experiments, our proposal might be extended with a large scale, and allows to simulate some phenomena of topological matters. We hope more potential applications can be found with our proposed method, include realizing nonreciprocal acoustic quantum devices and topological quantum simulations in phononic systems.

\begin{acknowledgments}
All the quantum dynamical simulations are based on open source code QuTiP~\cite{Johansson12qutip,Johansson13qutip}. The authors acknowledge fruitful discussions with Dr. Jie-Qiao Liao.
XW is supported by China Postdoctoral Science
Foundation No. 2018M631136, and the National Science Foundation of
China under Grant No. 11804270. HRL is supported by the National Science Foundation of China (Grant No.11774284).
\end{acknowledgments}


%

\end{document}